\documentclass{JHEP3}
\usepackage{amsmath}
\input epsf
\usepackage{epsfig}
\usepackage{amssymb}
\usepackage[active]{srcltx}

\newcommand{\bea}{\begin{eqnarray}}
\newcommand{\eea}{\end{eqnarray}}
\newcommand{\bean}{\begin{eqnarray*}}
\newcommand{\eean}{\end{eqnarray*}}
\newcommand{\nn}{\nonumber \\}

\def\W #1{\widetilde{#1}}

\def\braket#1{\left\langle #1 \right\rangle}

\def\gb #1{ \left\langle #1 \right]}

\def\a{{\alpha}}

\def\la{\lambda}
\def\eps{\epsilon}

\def\vev{\braket}

\def\bvev#1{\left[ #1 \right]}
\def\Spaa{\vev}
\def\Spbb{\bvev}
\def\Spab{\gb}

\def\Label#1{\label{#1}%
  \smash{\hbox to0pt{\raise1ex\hbox{\tiny[#1]}\hss}}}

\preprint{
\\
{\tt hep-th/}}
\title{A note on the boundary contribution with bad deformation
in gauge theory}
\author{ Bo Feng\footnote{email address: b.feng@cms.zju.edu.cn},
Chang-Yong Liu\footnote{email address: lcy@itp.ac.cn}\\{Center of
Mathematical Science, Zhejiang
 University, Hangzhou, China}}
\abstract{Motivated by recently progresses in the study of BCFW
recursion relation with nonzero boundary contributions for theories
with scalars and fermions\cite{Bofeng}, in this short note we
continue the study of boundary contributions of gauge theory with
the bad deformation. Unlike cases with scalars or fermions, it is
hard to use
 Feynman diagrams directly to obtain  boundary contributions, thus
 we propose another method based on the ${\cal N}=4$ SYM theory.
 Using this method, we are able to
 write down a  useful on-shell
recursion relation to calculate  boundary contributions from related
theories.  Our result shows the cut-constructibility of gauge theory
even with the bad deformation in some generalized sense.}

\begin{document}
\section{Motivations}
Although generally one can compute any scattering amplitude by
Feynman diagrams, as long as the theory has a Lagrangian
description, this method is usually complicated in practical
applications. In order to make  calculations of tree-level and
loop-level amplitudes more efficiently, many methods have been
suggested in  past few decades, among them there is the on-shell
BCFW recursion relation\cite{Britto:2004ap, Britto:2005fq}, which
was inspired by Witten's twistor program \cite{Witten:2003nn}. Using
the BCFW recursion relation, one can construct tree-level amplitudes
in terms of sub-amplitudes with fewer external particles, thus it
has reduced a big, difficult problem into several smaller and easier
ones. The on-shell recursion relation, together with the leading
singularities\cite{Britto:2004nc}, play important roles in  recent
developments of S-matrix
program\cite{ArkaniHamed:2009dn,ArkaniHamed:2009vw,ArkaniHamed:2009sx,
ArkaniHamed:2009dg,Boels:2010mj}.

The key idea of BCFW recursion relation is to pick up two special
momenta $p_i, p_j$ and do the following deformation (hence we will
call it "BCFW deformation") using an auxiliary momentum $q$:
\bea p_i(z)=p_1+z q,~~~~p_j(z)=p_j-z q~,
~~~~~\label{BCFW-deform-1}\eea
while  other momenta are untouched. With this deformation, the
momentum conservation is still kept. Furthermore, the momentum $q$
is chosen to satisfy conditions $q^2=0$, $p_i\cdot q=p_j\cdot q=0$
so that the deformed $p_i(z)$ and $p_j(z)$ are kept on-shell. It is
only possible for complex $q$ and space-time dimension $D\geq 4$.
With the deformed on-shell amplitude ${\cal A}(z)$ over single
complex variable $z$,  we consider following contour integration
\bea {\cal B} =\oint_C {{\cal A}(z)\over z}d z\eea
where contour $C$ is a big enough circle around $z=0$. We can
evaluate the integration by two different ways, either by contour
around $z=\infty$ or by contour of big circle around the origin.
Identified these two results we obtain
\bea {\cal A}(z=0)= -\sum_{z_\a} {\rm Res}\left( {{\cal A}(z)\over
z}\right)+{\cal B}~,~~~\label{BCFW-rel-1}\eea
where ${\cal A}(z=0)$ is the wanted physical amplitude and ${\cal
B}$ is the boundary contribution. Residues of poles in the right
hand side of (\ref{BCFW-rel-1}) can always be calculated using
factorization properties from lower-point on-shell amplitudes. In
other words, expression (\ref{BCFW-rel-1}) shows that for any
theory, some parts of tree amplitudes are always "on-shell
constructible". Contrast to that,  the boundary contribution ${\cal
B}$ is the obstacle for the application of  BCFW recursion relation.
If boundary contribution is zero under the chosen BCFW deformation,
the whole amplitude is cut-constructible as emphasized in
\cite{Paolo:2007}. But for some theories there are no such
deformations available to  makes ${\cal B}$ vanishing, thus we must
consider the boundary contribution  ${\cal B}$ when computing ${\cal
A}(z=0)$. So it is hard to see if the theory is cut-constructible or
not when ${\cal B}$ is not zero. Many theories we are familiar with
have zero boundary contributions, which usually can not be easily
inferred from Feynman diagrams and we should rely on other methods
as shown in \cite{ArkaniHamed:2008yf}-\cite{Cheung:2008dn}. Some
theories with non-zero boundary contributions have been discussed in
\cite{Bofeng}, and they are found to be cut-constructible.

For gauge theory, it is well known that one can always find good
deformation with zero boundary contribution ${\cal B}=0$
\cite{Britto:2004ap, Britto:2005fq}, such as deformations
$\Spab{i^+|j^-}$, $\Spab{i^+|j^+}$ and $\Spab{i^-|j^-}$ , i.e.,
\bea \la_i(z)=\la_i+z\la_j,~~~~
\W\la_j(z)=\W\la_j-z\W\la_i~.~~~~\label{Spab-deform}\eea
With these deformations one can calculate  tree-level amplitudes
${\cal A}$ efficiently. However, there is one bad deformation
$\Spab{i^-|j^+}$ which does not make the ${\cal A}(z)$ vanish at
infinity.
Although in practise one can always choose a good deformation to
simplify calculations and avoid the boundary contributions, it is
still important to study the non-zero boundary contributions from
the theoretical aspect. Especially we want to ask whether the theory
is still cut-constructible with nonzero boundary contributions, and
what is the physical implication of these non-zero values.

To calculate non-zero boundary contributions, there are many ways to
follow. The first method is to compute them from Feynman diagrams,
as have been done in paper \cite{Bofeng} for theories with scalars.
However, by some simple analysis it is easy to see that this method
is not very practical for gauge theory. Unlike theories with scalars
and fermions where only a few types of Feynman diagrams give nonzero
boundary contributions,  there are many  types of Feynman diagrams
giving potential nonzero boundary contributions in gauge theory, as
long as the line from $i$ to $j$ has no more than four four-point
vertexes. The second method  is to calculate the full amplitude with
good deformation and then deform it with bad deformation to get the
boundary contributions.  This method  can get the boundary
contributions straightforwardly, but it does not give us the insight
whether the boundary part is cut-constructible or not.

In this short note, we propose an alternative method to recursively
calculate nonzero boundary contributions  with bad deformation for
gauge theory and related theory, i.e., the theory with scalars and
fermions with ${\cal N}=4$ interaction terms. The note is organized
as following. In section two we will derive an on-shell recursion
relation for boundary contributions from the ${\cal N}=4$ SYM
theory. In section three we will illustrate our idea by two simple
examples. One of them is the gluon MHV amplitude and another one,
the NMHV amplitude with two fermions.  In section four we will give
a brief summary of our results.

\section{The general framework}

It is well known that tree-level amplitudes of pure gluons are
identical to these obtained from ${\cal N}=4$ super-Yang-Mills (SYM)
theory. In ${\cal N}=4$ SYM theory, one can group all components
into following on-shell superfield
\cite{Nair:1988bq,Witten:2003nn,ArkaniHamed:2008gz,Drummond}
\bea \Phi(p,\eta) & = & G^+(p)+\eta^A \psi_A^+ +{1\over 2}
\eta^A\eta^B S_{AB}+{1\over 3!}\eta^A\eta^B \eta^C \eps_{ABCD}
\psi^{D-}+{1\over 4!}\eta^A\eta^B \eta^C \eta^D\eps_{ABCD}
G^-(p)\eea
with Grassmann coordinate $\eta^A, A=1,2,3,4$. Using superfields,
amplitudes can be written as functions of $(\la_i,\W\la_i,
\eta_i^A)$. For example, the super-MHV amplitude is given by Nair's
formula \cite{Nair:1988bq} as
\bea {\cal A}_n(\la,\W\la, \eta) = {\delta^4(\sum_i \la_i
\W\la_i)\delta^8(\sum_i \la_i \eta_i^A)\over
\Spaa{1|2}\Spaa{2|3}...\Spaa{n|1}}~.~~~\label{SUSY-MHV}\eea
To obtain the corresponding scattering amplitudes with various
component configurations we just need to expand above expression as
the series of  $\eta^A$.

Having the super-amplitudes, we  need also the supersymmetric
version of BCFW recursion relation, which is given in
\cite{Brandhuber:2008pf,ArkaniHamed:2008gz}. The super-BCFW
deformation contains  the usual $\Spab{i|j}$-deformation for two
chosen momenta as given by (\ref{Spab-deform}), as well as the
deformation for $\eta$ variables as
\bea \eta_j^A(z)=\eta_j^A-z \eta_i^A~,~~~~~\label{eta-deform} \eea
so that the super-energy-momentum conservation is preserved, i.e.,
$\sum_t \la_t \eta_t^A$ is invariant under the full super-BCFW
deformation. With this in mind the supersymmetric version of BCFW
recursion relation can be written as
\bea {\cal A}= \sum_{L,R} \int d^4\eta_P M_L (\la_i(z_P),
\la_P(z_P),\W\la_P(z_P),\eta_P(z_P)) {1\over P^2} M_R
(\W\la_j(z_P),\eta_j(z_P),
\la_P(z_P),\W\la_P(z_P),\eta_P(z_P))~.~~~~~\label{SUSY-BCFW} \eea
If the deformed super-amplitude ${\cal A}^{SUSY}(z)$ approaches to
zero when $z$ goes to infinity\footnote{It will be interesting to
see if there is supersymmetric theory with nonzero boundary
contributions for any super-BCFW-deformation.}, there is no boundary
contribution. A nice property of $\mathcal{N}=4$ super-amplitudes is
that no matter what helicities of $(i,j)$ are, ${\cal A}^{{\cal
N}=4}(z)$ approaches to zero with $z\to \infty$, i.e., there is no
bad deformation at all for ${\cal N}=4$ SYM theory. Having this
fact, we expand the super-amplitude as the series of $\eta$ as
\bea {\cal A}_n = \sum_{a_1.. a_n=0}^4 {\cal A}_{a_1... a_n}
\prod_{i=1}^n\eta_i^{a_i}~~~\label{SUSY-exp}\eea
where ${\cal A}_{a_1... a_n}$ denotes amplitude of  field
configurations specified by $(a_1,a_2,...,a_n)$. Under the
deformation (\ref{Spab-deform}) and (\ref{eta-deform}) we have
\bea {\cal A}_n(z) = \sum_{a_1.. a_n=0}^4  {\cal A}_{a_1... a_n}(z)
\prod_{i=1}^n\eta_i(z)^{a_i}~,~~~\label{SUSY-exp-z}\eea
where $\eta_j(z)=\eta_j-z\eta_i$ and $\eta_k(z)=\eta_k$ for all the
$k\neq j$. ${\cal A}_{a_1... a_n}(z)$ is the super-amplitude of
specific field configuration $(a_1, a_2,...,a_n)$ after
$\Spab{i|j}$-deformation. However, it is more suitable to consider
the boundary behavior using the form (\ref{SUSY-exp}) since each
super-amplitude of specific configurations is independent in form
(\ref{SUSY-exp}). To re-write the deformed super-amplitude expansion
(\ref{SUSY-exp-z}) into the form (\ref{SUSY-exp}), we expand
$\eta_j(z)$ in (\ref{SUSY-exp-z}) as follows\footnote{Since it is
the expansion of Grassman variables, we need to be careful with the
sign when changing two variables, which we have not been very
careful in our arguments.}
\bean & & \sum_{a_1.. a_n=0}^4 {\cal A}_{a_1... a_n}(z)
\eta_i^{a_i}(\eta_j-z\eta_i)^{a_j}\prod_{k=1, k\neq i,j}^n
\eta_k^{a_k} \nn
& =& \sum_{a_1.. a_n=0}^4{\cal A}_{a_1... a_n}(z)
(\sum_{t=0}^{a_j}\a_t\eta_i^{a_i+t} (-z)^t
\eta_j^{a_j-t})\prod_{k=1, k\neq i,j}^n \eta_k^{a_k} \nn
& = & \sum_{a_1.. a_n=0}^4 \eta_i^{a_i} \eta_j^{a_j} \prod_{k=1,
k\neq i,j}^n \eta_k^{a_k}\left[\sum_{t\geq0, 4\geq a_i+t, a_j-t\geq
0} \a_t (-z)^t {\cal A}_{ a_i+t, a_j-t}(z)\right] \eean
where $\a_t$ represents the binomial coefficient.  After above
expansion, we can now explain our method. For the left hand side of
(\ref{SUSY-exp-z}) we have
\bea \oint dz {{\cal A}_n(z)\over z}=0~,\eea
and using the fact that each term in the form (\ref{SUSY-exp}) is
independent, we get immediately
\bea \oint {dz\over z} \sum_{t\geq 0,4\geq a_i+t, a_j-t\geq 0} \a_t
(-z)^t {\cal A}_{a_1,..., a_i+t,...,
a_j-t,...,a_n}(z)=0~.~~~\label{frame}\eea
Equations (\ref{frame}) gives relations among amplitudes whose field
configuration differences from each other only at the positions $i$
and $j$.  The boundary contributions of ${\cal A}_{a_i,a_j}(z)$
after $\Spab{i|j}$-deformation are calculated by the following
contour integration
\bea {\cal B}_{a_i,a_j} =\oint_C {{\cal A}_{a_i,a_j}(z)\over z}d
z\eea
where the contour is a big enough circle around $z=0$. Thus one can
use the relation (\ref{frame}) to calculate boundary contributions
of certain amplitude through  pole contributions of other
amplitudes. For example, the boundary contribution of ${\cal
A}_{a_i,a_j}(z)$ is given by
\bea {\cal B}_{a_i,a_j} =\oint_C {{\cal A}_{a_1,..., a_i,...,
a_j,...,a_n}(z)\over z}d z =-\oint {dz\over z} \sum_{t> 0,4\geq
a_i+t, a_j-t\geq 0} \a_t (-z)^t {\cal A}_{a_1,..., a_i+t,...,
a_j-t,...,a_n}(z)~, \eea
where the presence of positive powers of $z$ in the numerator makes
only  physical poles contribute to  boundary values. Similar
framework can be obtained for boundary contributions of theories
with fermions, for which  one example will be presented in next
section. A direct implication of our result is that even with a bad
deformation in gauge theory, boundary contributions are also
cut-constructible in some sense, i.e, they can be obtained from pole
contributions of some other related theories recursively.

\section{Examples}

In this section, we will use two simple examples to demonstrate our
idea. We will discuss gluon MHV amplitude in detail as the first
example and then briefly for  another example.

\subsection{The first example: gluon MHV amplitude}

Since the purpose of this note is to understand  boundary
contributions of bad deformation,  not to use it in practical
calculations, we will illustrate our idea simply by using MHV
amplitude of gluons, which is given by
\bea {\cal A}(s^-, n^-) & = & {\Spaa{s|n}^4\over
\Spaa{1|2}\Spaa{2|3}...\Spaa{n|1}}~.\eea
We take the bad deformation  $\Spab{s^-| t^+}$ with $t<s$. The
$z$-dependence can be written down explicitly as
\bean {\cal A}_{t=0, s=4}& =&  {\Spaa{s+z t|n}^4\over
\Spaa{1|2}\Spaa{2|3}... \Spaa{s-1|s+z t}\Spaa{s+z
t|s+1}...\Spaa{n|1}}~,\eean
where for later convenience we have explicitly written the power of
$\eta$-expansion at positions $t,s$ in the ${\cal N}=4$
super-amplitude. Splitting it into pole part of $z$ as well as
divergent parts of $z^a, a\geq 0$, we obtain following boundary
values, i.e., the coefficient of  $z^0$-term. If $t\neq s-1$ or
$t\neq s+1$, the highest power of $z$ is $z^2$ and the coefficient
of $z^0$ term is given by
\bea {\cal B}_{t=0,s=4} & = & {\Spaa{n|t}^2\over \Spaa{s-1|t}^3
\Spaa{t|s+1}^3} \left( 6 \Spaa{n|s}^2 \Spaa{s-1|t}^2 \Spaa{t|s+1}^2
\right. \nn & & \left. -4
\Spaa{n|s}\Spaa{n|t}\Spaa{s-1|t}\Spaa{t|s+1}(\Spaa{s|s+1}\Spaa{s-1|t}+\Spaa{s-1|s}
\Spaa{t|s+1}) \right. \nn & & \left. +\Spaa{n|t}^2 (
\Spaa{s|s+1}^2\Spaa{s-1|t}^2+\Spaa{s|s+1}\Spaa{s-1|t}\Spaa{s|s-1}\Spaa{s+1|t}
+ \Spaa{s|s-1}^2\Spaa{s+1|t}^2)\right)~.~~\nn
~~~~\label{B-res-1}\eea
If $t=s-1$ or $t=s+1$, the highest power becomes $z^3$, and the
$z^0$ term would be (we just write down the case $t=s-1$)
\bea {\cal B}_{t=0,s=4} & = & {-\Spaa{n|s-1}\over
\Spaa{s|s-1}\Spaa{s-1|s+1}^4}\left( -\Spaa{n|s-1}^3
\Spaa{s|s+1}^3-6\Spaa{n|s}^2\Spaa{s-1|s+1}^2\Spaa{n|s-1}\Spaa{s|s+1}\right.\nn
& & \left. +4 \Spaa{n|s}\Spaa{s-1|s+1}(\Spaa{n|s-1}^2\Spaa{s|s+1}^2+
\Spaa{n|s}^2\Spaa{s-1|s+1}^2)\right) ~.~~~~\label{B-res-2}~~\eea
It is worth to notice that the numerical coefficients of each term
in expression (\ref{B-res-1}) and (\ref{B-res-2}) is $1,4,6,4,1$,
which are an indication of matter contents of ${\cal N}=4$ theory.

Now we apply our framework (\ref{frame}) to this example. The
relevant combination can be easily found to be
\bea \oint {dz\over z} \left( {\cal A}_{t=0,s=4}(z)- 4z {\cal
A}_{t=1,s=3}(z)+ 6 z^2 {\cal A}_{t=2,s=2}(z)- 4 z^3 {\cal
A}_{t=3,s=1}(z)+z^4 {\cal
A}_{t=4,s=0}(z)\right)=0~.~~~~~~\label{MHV-b-1} ~\eea
By using  following results
\bea {\cal A}_{t=4,s=0} & = & {\Spaa{t|n}^4\over
\Spaa{1|2}\Spaa{2|3}...\Spaa{n|1}},~~~~{\cal A}_{t=3,s=1} =
{\Spaa{s|n} \Spaa{t|n}^3 \over
\Spaa{1|2}\Spaa{2|3}...\Spaa{n|1}}~,\\
{\cal A}_{t=2,s=2}& = & {\Spaa{s|n}^2 \Spaa{t|n}^2 \over
\Spaa{1|2}\Spaa{2|3}...\Spaa{n|1}},~~~~{\cal A}_{t=1,s=3}=
{\Spaa{s|n}^3 \Spaa{t|n} \over
\Spaa{1|2}\Spaa{2|3}...\Spaa{n|1}}~,\eea
the integrand of (\ref{MHV-b-1}) becomes
\bean  {\Spaa{s|n}^4\over \Spaa{1|2}\Spaa{2|3}...\Spaa{s-1|s+z t}
\Spaa{s+z t|s+1}...\Spaa{n|1}}~,\eean
which gives zero boundary contribution as it should be. Then we
rewrite (\ref{MHV-b-1}) as follows
\bea  {\cal B}_{t=0,s=4} & \equiv & \oint {dz\over z}{\cal
A}_{t=0,s=4}(z) \nn
& = & - \oint {dz\over z} \left(- 4z {\cal A}_{t=1,s=3}(z)+ 6 z^2
{\cal A}_{t=2,s=2}(z)- 4 z^3 {\cal A}_{t=3,s=1}(z)+z^4 {\cal
A}_{t=4,s=0}(z)\right)~.~~~~\label{Exp-B} \eea
It is easy to see that one can use (\ref{Exp-B}) to calculate
boundary contributions in the first line. Since the $z$-factor
appears in numerator, there is no pole contribution at $z=0$ in
(\ref{Exp-B}), and all  contributions come from physical poles. To
calculate these terms we can use formula
\bea  \oint {dz A(z)\over z} z^k = -\sum_\a A_L(z_\a) {z_\a^k\over
P^2} A_R(z_\a),~~~k\geq 1~.~~~\label{Zk-rec}\eea
We will apply formula (\ref{Zk-rec}) to (\ref{Exp-B}) and compare
results with the one given by (\ref{B-res-1}) and (\ref{B-res-2}).
\begin{figure}
\center
\includegraphics[width=0.6\textwidth]{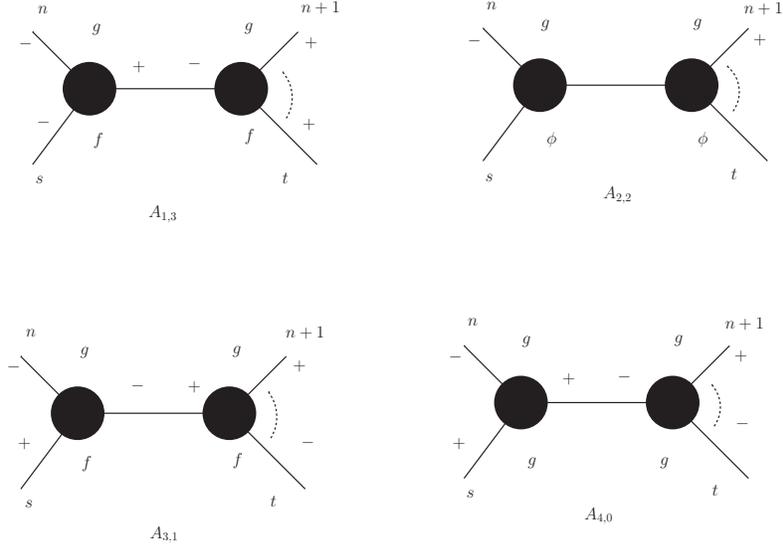}
\caption{\label{1} The four diagrams are these boundary
contributions for $A_{0,4}$ with $t=s-1$ and $n=s+1$ using the
recursion relation.}
\end{figure}
~\\

{\bf The case $t=s-1,n=s+1$:}

As a warm up let us start from the simplest case $t=s-1$ and
$n=s+1$. There are four terms from recursion relation (\ref{Exp-B}),
as illustrated in figure (\ref{1}). The first three diagrams do not
contribute because the left hand side of each diagram is three-point
MHV amplitude with all $\la$ proportional to each other, thus gives
zero value. So  boundary contributions of $A_{t=0,s=4}$ are given by
$A_L^{4,0}(z_\a) {z_\a^4\over P^2} A_R^{4,0}(z_\a)$. Using
 \bea
z_\a=-\frac{\Spaa{n| s}}{\Spaa{n| t}}~,~~ \eea
we obtain
 \bea {\cal
B}_{t=0,s=4} & = & A_L^{4,0}(z_\a) {z_\a^4\over P^2}
A_R^{4,0}(z_\a)=(-\frac{\Spaa{n| s}}{\Spaa{n| t}})^4 A_L^{4,0}(z_\a)
{1\over P^2} A_R^{4,0}(z_\a)={\Spaa{s|n}^4\over
\Spaa{1|2}\Spaa{2|3}...\Spaa{n|1}}=A_{0,4}(0)  \eea
which agrees with the one given by (\ref{B-res-2}).

\begin{figure}
\center
\includegraphics[width=0.6\textwidth]{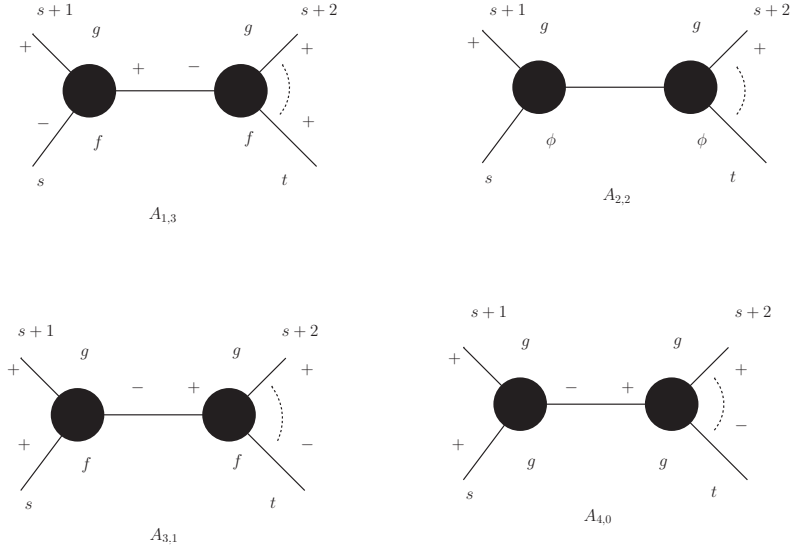}
\caption{\label{2} The four diagrams are these boundary
contributions for $A_{0,4}$ with $t=s-1$ and $n \neq s+1$ using the
recursion relation.}
\end{figure}
~\\

{\bf The case $t=s-1, n\neq s+1$:}

The second case  we are considering is $t=s-1$ and $n\neq s+1$.
Again there are four terms from recursion relation (\ref{Exp-B}), as
shown in figure (\ref{2}). All these four diagrams give nonzero
boundary contributions to $A_{t=0,s=4}$. By using
\bea \label{z1} z_\a=-\frac{\Spaa{ s+1| s}}{\Spaa{ s+1| t}}~,~~ \eea
we have
\bea \label{s} \mid s(z_\a)\rangle=\mid \hat{s}\rangle=\mid
s\rangle-\frac{\Spaa{s+1| s}}{\Spaa{s+1| t}} \mid
t\rangle=-\frac{\Spaa{t|s}}{\Spaa{ s+1|t}}\mid s+1\rangle \eea
where we have used the Schouten identity. All contributions from
four diagrams in figure (\ref{2}) have common denominator as
\bea & &  P(0)^2\Spaa{\hat{t}|\hat{P}}\Spaa{\hat{P}|s+2}\Spaa{s+2|
s+3}\cdots \Spaa{t-1|\hat{t}}\Spbb{\hat{s}|s+1}\Spbb{s+1|
-\hat{P}}\Spbb{-\hat{P}|\hat{s}} \nn &= &
\Spbb{s+1|s}^4\Spaa{s|s+1}\Spaa{s+1|s+2}\cdots \Spaa{t-1| t}\Spaa{
t|s} \eea
and these four numerators are given separately as
\bea \label{sabc} A_{1,3}: & ~~~&
-4\Spbb{\hat{s}|s+1}\Spbb{\hat{P}|s+1}^3\Spaa{\hat{P}|n}^3
\Spaa{\hat{t}|n}z_\a
=4z_\a\Spbb{\hat{s}|s+1}^4\Spaa{n|\hat{s}}^3\Spaa{\hat{t}|n},\nn
A_{2,2}: & ~~~&
\Spbb{\hat{s}|s+1}^2\Spbb{\hat{P}|s+1}^2\Spaa{\hat{P}|n}^2
\Spaa{\hat{t}|n}^26z_\a^2
=6z_\a^2\Spbb{\hat{s}|s+1}^4\Spaa{n|\hat{s}}^2\Spaa{\hat{t}|n}^2\nn
A_{3,1}: & ~~~&
-4z_\a^3\Spbb{\hat{s}|s+1}^3\Spbb{\hat{P}|s+1}^1\Spaa{\hat{P}|n}^1
\Spaa{\hat{t}|n}^3
=4z_\a^3\Spbb{\hat{s}|s+1}^4\Spaa{n|\hat{s}}^1\Spaa{\hat{t}|n}^3\nn
A_{4,0}: &~~~& z_\a^4\Spbb{\hat{s}|s+1}^4\Spaa{\hat{t}|n}^4\eea
Sum up these four numerators we have
\bea
\Spbb{\hat{s}|s+1}^4[\Spaa{n|\hat{s}}+z_\a\Spaa{\hat{t}|n}]^4-\Spbb{\hat{s}|s+1}^4
\Spaa{n|\hat{s}}^4~.~~\eea
Combining with the common denominator we get the final result
 \bea \label{zwe}
{\cal B}_{t=0,s=4}& = &
\frac{[\Spaa{n|\hat{s}}+z_\a\Spaa{\hat{t}|n}]^4-\Spaa{n|\hat{s}}^4}{\Spaa{s|s+1}\Spaa{s+1|s+2}\cdots
\Spaa{t-1| t}\Spaa{ t|s}}\nn
& =
&\frac{\Spaa{n|s}^4-[\frac{\Spaa{n|s+1}\Spaa{t|s}}{\Spaa{t|s+1}}]^4}{\Spaa{s|s+1}\Spaa{s+1|s+2}\cdots
\Spaa{t-1| t}\Spaa{ t|s}}  \eea
where in the second step we have used (\ref{z1}) and (\ref{s}). This
result agrees with the one given by (\ref{B-res-2}).

\begin{figure}
\center
\includegraphics[width=0.6\textwidth]{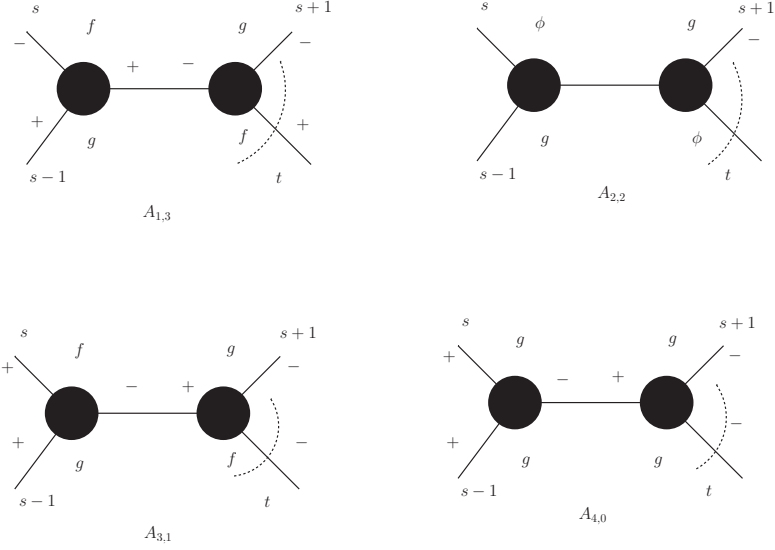}
\caption{\label{3} These four diagrams are the A part of  boundary
contributions for $A_{0,4}$ with $t\neq s-1$, $t\neq s+1$ and
$n=s+1$ using the recursion relation.}
\end{figure}
~\\

{\bf Case $t\neq s-1$, $t\neq s+1$ and n=s+1:}

\begin{figure}
\center
\includegraphics[width=0.2\textwidth]{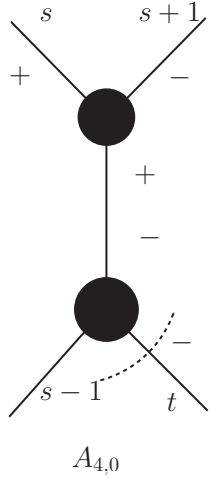}
\caption{\label{4} The diagram is the B part of  boundary
contributions for $A_{0,4}$ with $t\neq s-1$, $t\neq s+1$ and
$n=s+1$ using the recursion relation.}
\end{figure}

The third case  we are considering is  $t\neq s-1$, $t\neq s+1$ and
$n=s+1$. There are five diagrams contributing to  boundary values.
We divide these five diagrams into two parts, as part A shown in
figure (\ref{3}) and part B, in figure (\ref{4}). We first calculate
contributions of part A from figure (\ref{3}), where $z_\a^A$ is
given by
\bea \label{z} z_\a^A=-\frac{\Spaa{ s-1| s}}{\Spaa{ s-1| t}}~.~~
\eea
The common denominator of these four contributions is
\bea & &P(0)^2\Spaa{\hat{t}|\hat{P}}\Spaa{\hat{P}|s+1}\Spaa{s+1|
s+2}\cdots \Spaa{t-1|\hat{t}}\Spbb{s-1|\hat{s}}\Spbb{\hat{s}|
\hat{P}}\Spbb{\hat{P}|s-1}\nn
&=&\Spbb{s-1|s}^4\Spaa{t|t+1}\cdots\Spaa{s-1|s}\Spaa{\hat{s}|
s+1}\Spaa{s+1|s+2}\cdots\Spaa{t-1|t}~.~~ \eea
By using Schouten identity, we have
\bea \Spaa{\hat{s}| s+1}=\frac{\Spaa{s|t}\Spaa{s-1| s+1}}{\Spaa{s-1|
t}} ~,~~\eea
thus the denominator becomes
\bea \Spbb{s-1|s}^4\frac{\Spaa{s|t}\Spaa{s-1| s+1}}{\Spaa{s-1|
t}\Spaa{s|s+1}}\prod_i\Spaa{i|i+1}~.~~  \eea
By similar calculations for these four  numerators as in
(\ref{sabc}) we get
\bea &
&\Spbb{\hat{s}|s-1}^4[\Spaa{n|\hat{s}}+z_\a\Spaa{t|n}]^4-\Spbb{\hat{s}|s-1}^4
\Spaa{n|\hat{s}}^4 \nn
&=&\Spbb{\hat{s}|s-1}^4\Spaa{n|s}^4-\Spbb{\hat{s}|s-1}^4
[\frac{\Spaa{n|s-1}\Spaa{s|t}}{\Spaa{s-1|t}}]^4 ~.~~\eea
Putting these results together we have part A contributions from
figure (\ref{3})
\bea \frac{\Spaa{s-1| t}\Spaa{s|s+1}}{\Spaa{s|t}\Spaa{s-1|
s+1}}\frac{1}{\prod_i\Spaa{i|i+1}}[\Spaa{n|s}^4-[\frac{\Spaa{n|s-1}\Spaa{s|t}}{\Spaa{s-1|t}}]^4]~.~~
\eea
The part B contribution from figure (\ref{4}) can be calculated by
similar way and the result is
\bea \frac{\Spaa{s+1| t}\Spaa{s-1|s}}{\Spaa{s-1|s+1}\Spaa{s|
t}}\frac{1}{\prod_i\Spaa{i|i+1}}\Spaa{n|s}^4 ~.~~\eea
By using Schouten identity $ \Spaa{s-1| t}\Spaa{s|s+1}+\Spaa{s+1|
t}\Spaa{s-1|s}=\Spaa{s|t}\Spaa{s-1| s+1}$,
we can combine part A and part B contributions  together and get the
final result for  boundary values
$A_{0,4}$ with $t\neq s-1$, $t\neq s+1$ and n=s+1 is \bea {\cal
B}_{t=0,s=4}=\frac{1}{\prod_i\Spaa{i|i+1}}[\Spaa{n|s}^4-
\frac{\Spaa{s-1| t}\Spaa{s|s+1}}{\Spaa{s|t}\Spaa{s-1|
s+1}}[\frac{\Spaa{n|s-1}\Spaa{s|t}}{\Spaa{s-1|t}}]^4]~.~~ \eea
The result agrees with the one given by (\ref{B-res-1}).
~\\

{\bf The case $t\neq s-1$, $t\neq s+1$, $n\neq s+1$ and $n\neq
s-1$:}

\begin{figure}
\center
\includegraphics[width=0.6\textwidth]{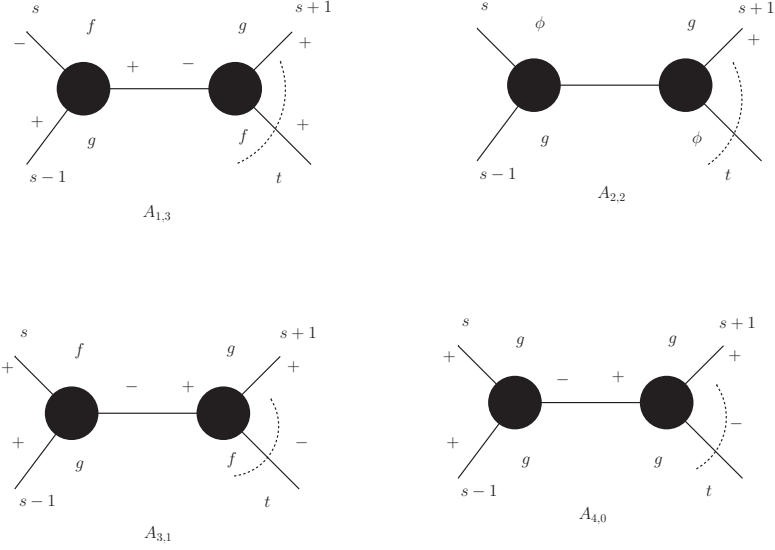}
\caption{\label{5} These diagrams are the A part of the boundary
contributions for $A_{0,4}$ with $t\neq s-1$, $t\neq s+1$, $n\neq
s+1$ and $n\neq s-1$ using the recursion relation.}
\end{figure}

\begin{figure}
\center
\includegraphics[width=0.5\textwidth]{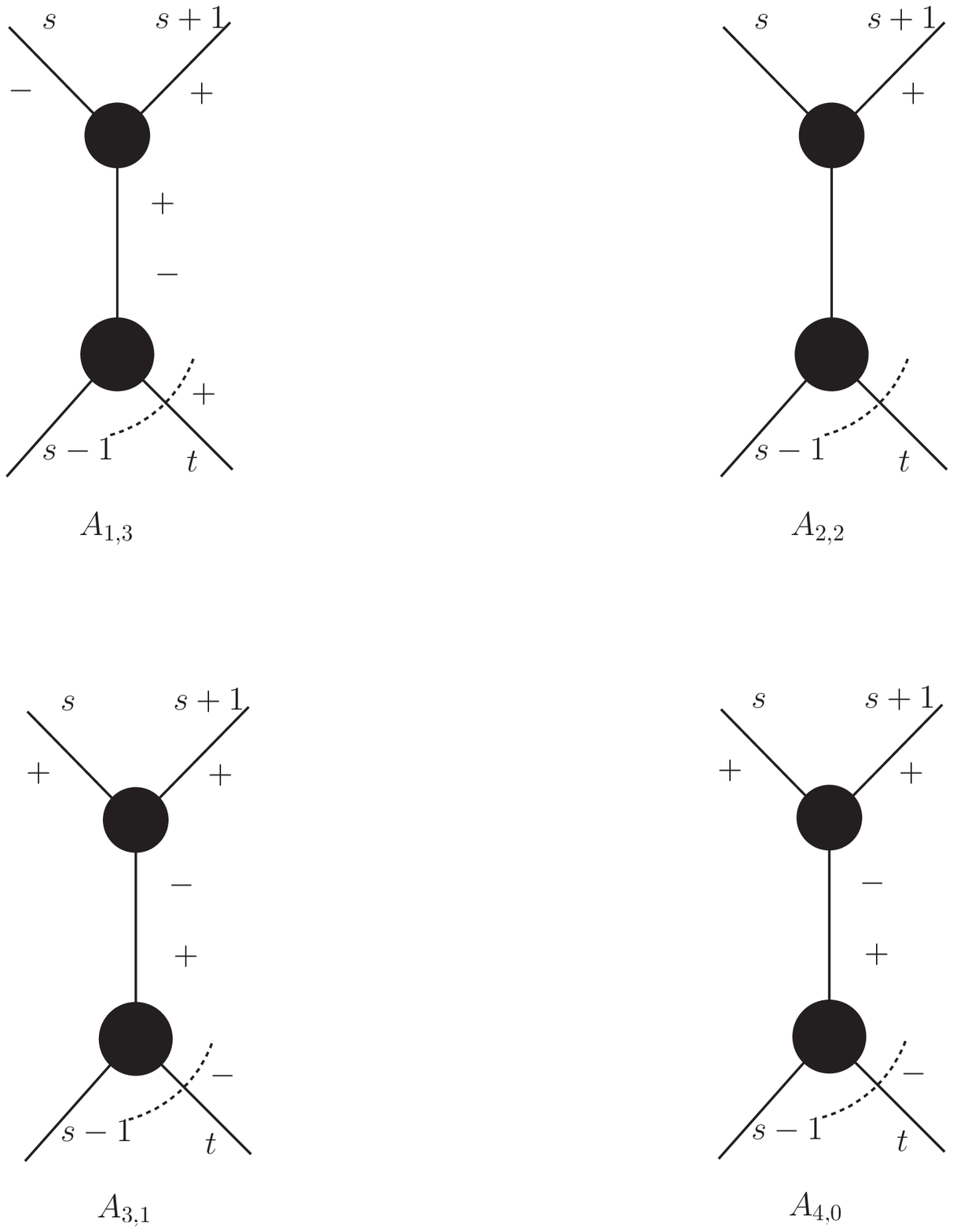}
\caption{\label{6} These diagrams are the B part of the boundary
contributions for $A_{0,4}$ with $t\neq s-1$, $t\neq s+1$, $n\neq
s+1$ and $n\neq s-1$ using the recursion relation.}
\end{figure}

The last case  we are  considering is  $t\neq s-1$, $t\neq s+1$,
$n\neq s+1$ and $n\neq s-1$. There are totally eight diagrams
contributing to $A_{0,4}$, again we  divide  eight diagrams into two
parts, as part A shown in figure (\ref{5}) and part B, in figure
(\ref{6}).

By similar calculations as done in (\ref{zwe}) we get  results for
part A in figure (\ref{5}) as
\bea \frac{\Spaa{s-1| t}\Spaa{s|s+1}}{\Spaa{s|t}\Spaa{s-1|
s+1}}\frac{1}{\prod_i\Spaa{i|i+1}}
[\Spaa{n|s}^4-[\frac{\Spaa{n|s-1}\Spaa{s|t}}{\Spaa{s-1|t}}]^4] \eea
and similarly the contribution for part B in figure (\ref{6}) as
\bea \frac{\Spaa{s+1| t}\Spaa{s-1|s}}{\Spaa{s-1|s+1}\Spaa{s|
t}}\frac{1}{\prod_i\Spaa{i|i+1}}[\Spaa{n|s}^4-[\frac{\Spaa{n|s+1}\Spaa{t|s}}{\Spaa{t|s+1}}]^4]
~.~~\eea

So the final boundary value for $A_{0,4}$ with $t\neq s-1$, $t\neq
s+1$, $n\neq s+1$ and $n\neq s-1$ is
\bea {\cal
B}_{t=0,s=4}&=&\frac{1}{\prod_i\Spaa{i|i+1}}[\Spaa{n|s}^4-
\frac{\Spaa{s-1| t}\Spaa{s|s+1}}{\Spaa{s|t}\Spaa{s-1|
s+1}}[\frac{\Spaa{n|s-1}\Spaa{s|t}}{\Spaa{s-1|t}}]^4 \nn
&-&\frac{\Spaa{s+1| t}\Spaa{s-1|s}}{\Spaa{s-1|s+1}\Spaa{s|
t}}[\frac{\Spaa{n|s+1}\Spaa{t|s}}{\Spaa{t|s+1}}]^4] ~.~~ \eea
The result agrees with the one given by (\ref{B-res-1}).

\subsection{The second example: amplitude with two fermions and four gluons $\Lambda_1^+\Lambda_2^-g_3^-g_4^-
g_5^+g_6^+$ }

%
\begin{figure}
\center
\includegraphics[width=0.6\textwidth]{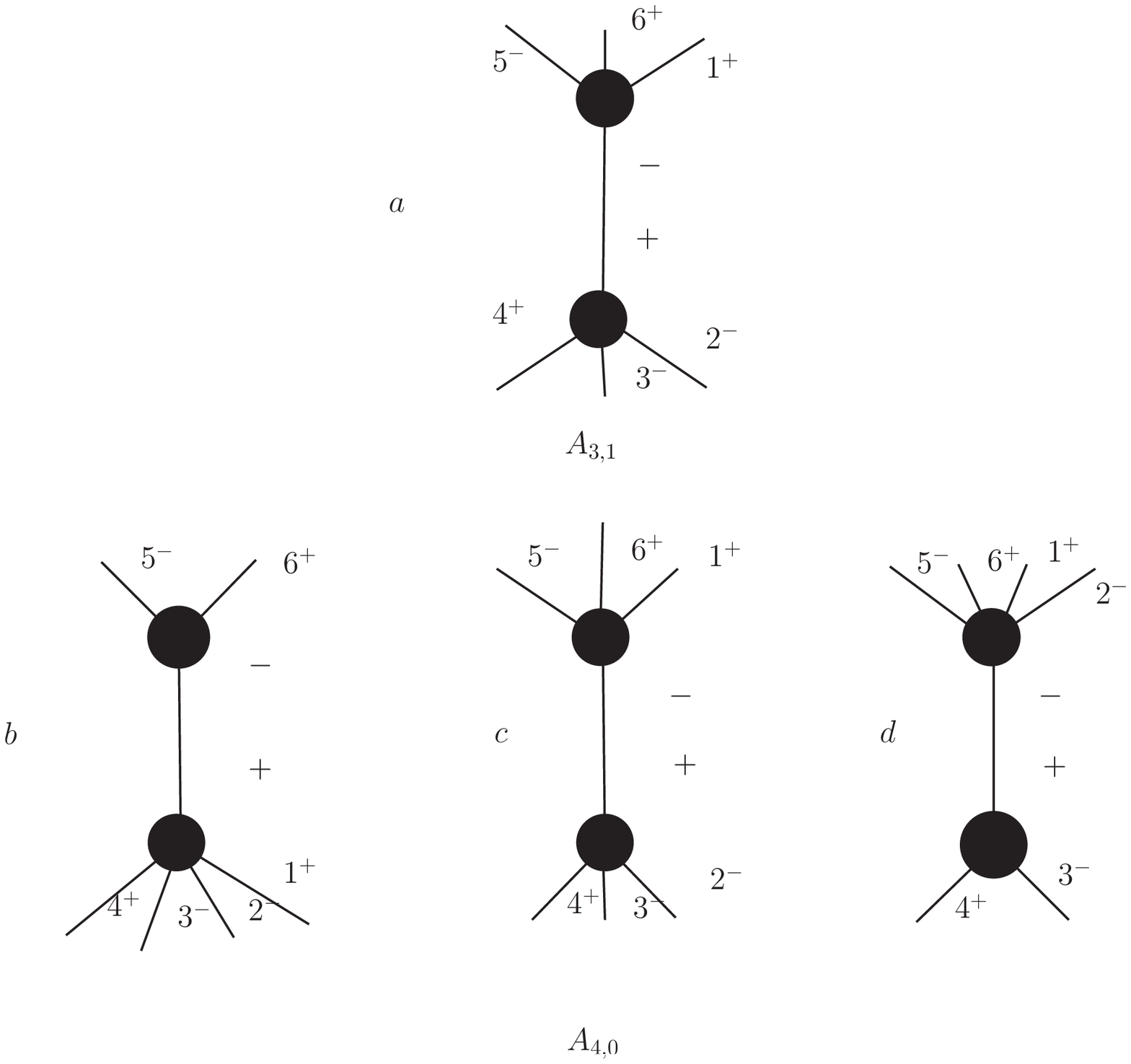}
\caption{\label{7} The boundary term for
$A(\Lambda_1^+\Lambda_2^-g_3^-g_4^- g_5^+g_6^+)$  with
$\Spab{4^-|5^+}$ shifting using the recursion relation.}
\end{figure}

The $A(\Lambda_1^+\Lambda_2^-g_3^-g_4^- g_5^+g_6^+)$ amplitude has
been calculated in \cite{Luo} by choosing a good deformation, here
we will discuss  boundary contributions with bad deformation for
this one, thus provide a generalization of our discussions to cases
besides gluons. We will consider two different bad deformations.

First we consider the bad shifting $\Spab{4^-|5^+}$. With this
choice, there is no contribution from physical poles and the whole
amplitude is given by boundary contributions. There are four
diagrams under $\Spab{4^-|5^+}$ deformation, as shown in figure
(\ref{7}). The recursion relation for boundary contributions is
given by
\bea   {\cal B}_{t=0,s=4} & \equiv & \oint {dz\over z}{\cal
A}_{t=0,s=4}(z) \nn
& = & - \oint {dz\over z} \left(-  z^3 {\cal A}_{t=3,s=1}(z)+z^4
{\cal A}_{t=4,s=0}(z)\right)~.~~~~ \eea
according to the general formula (\ref{frame}). Compared to the one
given by (\ref{Exp-B}), we see that there are only two terms with
numerator coefficient $1$. The reason is that this amplitude has
fermions $\Lambda_1,\Lambda_2$, so in order to get non-zero
contributions, $\Lambda$ must pair up with the same kind of fermions
(recall that there are four different fermions in ${\cal N}=4$ SYM
theory).

The boundary contributions from figure (\ref{7}.a) and (\ref{7}.c)
are given by
\bea & &-A_L^{3,1}(z_a) {z_a^3\over
P^2} A_R^{3,1}(z_a)+A_L^{4,0}(z_c) {z_c^4\over P^2}
A_R^{4,0}(z_c)\nn
 &= &
-\frac{s_{561}^2\Spab{1|5+6|2}}{\Spaa{5|6}\Spaa{1|6}
\Spbb{2|3}\Spbb{3|4}\Spab{5|1+6|2}\Spab{1|5+6|4}} \eea
while the boundary contributions from figure (\ref{7}.b) and figure
(\ref{7}.d) are given by
\bea -\frac{\Spaa{3|2}^2\Spaa{3|1}\Spbb{5|6}^3}
{s_{456}\Spaa{1|2}\Spbb{4|5}\Spab{3|4+5|6}\Spab{1|5+6|4}}
-\frac{\Spbb{6|1}^2\Spbb{6|2}\Spaa{3|4}^3}{s_{345}
\Spaa{4|5}\Spbb{1|2}\Spab{3|1+2|6}\Spab{5|1+6|2}}\eea
where $s_{561}=(k_5+k_6+k1)^2, s_{456}=(k_4+k_5+k_6)^2$ and
$s_{345}=(k_3+k_5+k_5)^2 $. Putting these four terms together, we
find that
  \bea \label{2g} {\cal
B}_{t=0,s=4}&= &
-\frac{s_{561}^2\Spab{1|5+6|2}}{\Spaa{5|6}\Spaa{1|6}
\Spbb{2|3}\Spbb{3|4}\Spab{5|1+6|2}\Spab{1|5+6|4}}\nn
&-&\frac{\Spaa{3|2}^2\Spaa{3|1}\Spbb{5|6}^3}
{s_{456}\Spaa{1|2}\Spbb{4|5}\Spab{3|4+5|6}\Spab{1|5+6|4}}
-\frac{\Spbb{6|1}^2\Spbb{6|2}\Spaa{3|4}^3}{s_{345}
\Spaa{4|5}\Spbb{1|2}\Spab{3|1+2|6}\Spab{5|1+6|2}}\nn
&=&A(\Lambda_1^+\Lambda_2^-g_3^-g_4^- g_5^+g_6^+)~,
~~~~\label{fermion-b}\eea
where  the last identity can be checked with the result obtained in
\cite{Luo}.

Our general framework can also be used to solve the problem raised
in \cite{Luo1}, where the author concluded that one can not use
nearby fermions to take BCFW deformation if the helicity
configuration is $(+,-)$ or $(-,+)$. The author also stated that the
pair of one fermion and an adjacent gluon with the same helicity is
also not suitable for BCFW deformation. Now we understand that one
can use nearby fermions to do BCFW deformation if one is able to
find  boundary contributions. Using the recursion relation, the
boundary contributions with $\Spab{1^+|2^-}$ shifting are given by
\bea {\cal B}'&= & \oint {dz\over z}  A_{1,3}(z)= \oint {dz\over z}
z^3 A_{4,0}(z) =- A_L^{4,0}(z_\a) {z_\a^3\over P^2}
A_R^{4,0}(z_\a)~. ~~~~ \eea
Contributions from physical poles under this deformation is zero and
the whole amplitude is again given by boundary contributions.  There
are three diagrams with pole contributions from $A_{4,0}$, i.e., the
pure gluon helicity configuration $(-,+,-,-,+,+)$. These three terms
are exactly the one given by (\ref{fermion-b}).

\section{Conclusion and discussions }
  In this paper we have studied
boundary contributions with bad deformation in gauge theory.  We
deduce a very useful on-shell recursion relation to calculate the
boundary contributions from ${\cal N}=4$ SUSY amplitudes. It
provides useful understanding of bad deformations  although we can
always choose good deformations to calculate the gauge theory
amplitudes. Especially our recursion relation shows the
cut-constructibility of boundary contributions in generalized sense,
i.e., they are given by  pole contributions of related theory.
Obviously same consideration can be done for amplitudes with
gravitons.

\section*{Acknowledgments}

BF is supported by fund from Qiu-Shi, the Fundamental Research Funds
for the Central Universities with contract number 2009QNA3015, as
well as Chinese NSF funding under contract No.10875104.



\begin{thebibliography}{999}
\bibitem{Bofeng}
  Bo Feng, Junqi Wang, Yihong Wang and Zhibai Zhang,
  ``BCFW Recursion Relation with Nonzero Boundary Contribution,''
  JHEP {\bf 1001}, 019 (2010)
  [arXiv:0911.0301 [hep-th]].



\bibitem{Britto:2004ap}
  R.~Britto, F.~Cachazo and B.~Feng,
  Nucl.\ Phys.\  B {\bf 715}, 499 (2005)
  [arXiv:hep-th/0412308].

\bibitem{Britto:2005fq}
  R.~Britto, F.~Cachazo, B.~Feng and E.~Witten,
  Phys.\ Rev.\ Lett.\  {\bf 94}, 181602 (2005)
  [arXiv:hep-th/0501052].

\bibitem{Witten:2003nn}
  E.~Witten,
  Commun.\ Math.\ Phys.\  {\bf 252}, 189 (2004)
  [arXiv:hep-th/0312171].

\bibitem{Britto:2004nc}
  R.~Britto, F.~Cachazo and B.~Feng,
  Nucl.\ Phys.\  B {\bf 725}, 275 (2005)
  [arXiv:hep-th/0412103].


\bibitem{ArkaniHamed:2009dn}
  N.~Arkani-Hamed, F.~Cachazo, C.~Cheung and J.~Kaplan,
  JHEP {\bf 1003}, 020 (2010)
  [arXiv:0907.5418 [hep-th]].

\bibitem{ArkaniHamed:2009vw}
  N.~Arkani-Hamed, F.~Cachazo and C.~Cheung,
  JHEP {\bf 1003}, 036 (2010)
  [arXiv:0909.0483 [hep-th]].

\bibitem{ArkaniHamed:2009sx}
  N.~Arkani-Hamed, J.~Bourjaily, F.~Cachazo and J.~Trnka,
  arXiv:0912.3249 [hep-th].

\bibitem{ArkaniHamed:2009dg}
  N.~Arkani-Hamed, J.~Bourjaily, F.~Cachazo and J.~Trnka,
  arXiv:0912.4912 [hep-th].

\bibitem{Boels:2010mj}
  R.~H.~Boels,
  arXiv:1003.2989 [hep-th].

\bibitem{Paolo:2007}
  Paolo.~Benincasa,   Freddy.~Cachazo,
  ``Consistancy Conditions On The S-Matix Of Massless Particles''
  arXiv:0705.4305[hep-th].

\bibitem{ArkaniHamed:2008yf}
  N.~Arkani-Hamed and J.~Kaplan,
  JHEP {\bf 0804}, 076 (2008)
  [arXiv:0801.2385 [hep-th]].




\bibitem{Vaman:2005dt}
  D.~Vaman and Y.~P.~Yao,
  JHEP {\bf 0604}, 030 (2006)
  [arXiv:hep-th/0512031].

\bibitem{Draggiotis:2005wq}
  P.~D.~Draggiotis, R.~H.~P.~Kleiss, A.~Lazopoulos and C.~G.~Papadopoulos,
  Eur.\ Phys.\ J.\  C {\bf 46}, 741 (2006)
  [arXiv:hep-ph/0511288].

\bibitem{Benincasa:2007qj}
  P.~Benincasa, C.~Boucher-Veronneau and F.~Cachazo,
  JHEP {\bf 0711}, 057 (2007)
  [arXiv:hep-th/0702032].



\bibitem{Cheung:2008dn}
  C.~Cheung,
  arXiv:0808.0504 [hep-th].

\bibitem{Nair:1988bq}
  V.~P.~Nair,
  Phys.\ Lett.\  B {\bf 214}, 215 (1988).



\bibitem{ArkaniHamed:2008gz}
  N.~Arkani-Hamed, F.~Cachazo and J.~Kaplan,
  arXiv:0808.1446 [hep-th].


\bibitem{Drummond}
 J.M.Drummond, J.Henn, G.P.Korchemsky and E.Sokatchev,
 ``Dual superconformal symmetry of scattering amplitudes in
  $\mathcal {N}$=4 super-Yang Mills theory,''
  Nucl.\ Phys.\  B {\bf 828}(2010)317
  [arXiv:0807.1095 [hep-th]].

\bibitem{Brandhuber:2008pf}
  A.~Brandhuber, P.~Heslop and G.~Travaglini,
  Phys.\ Rev.\  D {\bf 78}, 125005 (2008)
  [arXiv:0807.4097 [hep-th]].




\bibitem{Luo}
  Ming-xing Luo and Congkao Wen,
  ``Recursion relations for Tree Amplitudes in Super Gauge Theories,''
  JHEP {\bf 0503}, 004 (2005)
  [arXiv:hep-th/0501121].

\bibitem{Luo1}
  Ming-xing Luo and Congkao Wen,
  ``Compact Formulas for Tree Amplitudes of Six Partions,''
  Phys.Rev.D {\bf71}: 091501 (2005)
  [arXiv:hep-th/0502009].








\end{thebibliography}
\end{document}